\documentclass[11pt,a4paper]{article}
\pdfoutput=1

\usepackage[margin=2.8cm,bottom=3.5cm]{geometry}
\usepackage{cite}

\usepackage{amsmath, amssymb}
\usepackage{graphicx}
\usepackage{color}

\title{A $\phi^6$ soliton with a long-range tail}
\author{Andr\'e Amado\thanks{andre.amado@ufpe.br}\ \ and Azadeh Mohammadi\thanks{azadeh.mohammadi@df.ufpe.br}\\~\\Departamento de F\'isica, Universidade Federal de Pernambuco\\ 52171-900 Recife, Pernambuco, Brasil}

\date{\today}

\begin{document}

\maketitle
\begin{abstract}
We propose an analytically solvable sextic potential model with non-trivial soliton solutions connecting the trivial vacua. The model does not respect parity symmetry, and like $\phi^4$ theory has two minima. The soliton solutions and the consequent results are obtained in terms of the Lambert W function, i.e., the inverse function of $f(W) = We^W$. They have power-law asymptotics at one spatial infinity and exponential asymptotics at the other. We compare the solution with the kink of $\phi^4$ theory, which preserves the parity symmetry and has exponential asymptotics at both spatial infinities. Moreover, we study the full spectrum (bound and continuum states) of boson and fermion fields in the presence of the proposed soliton. We consider two types of coupling for the boson-soliton interaction and Yukawa coupling for the fermion-soliton interaction. Most results are derived analytically. This property renders the model a fertile ground for further study, including parity breaking related phenomena and long-range soliton-soliton interactions.
\end{abstract}

\section{Introduction}
There is a small class of nonlinear differential equations with soliton solutions. A soliton is a stable solution with localized energy density. It arises as a consequence of the interaction between nonlinearity and dispersion when a nonlinear sharpening term counterbalances the dispersive term. The competition between these two contributions shapes the structure of the soliton and provides its stability. In the language of topology, the soliton configuration has an associated conserved topological charge or winding number, which protects it against decay into a trivial configuration. Solitons are fascinating due to their mathematical properties. However, their usefulness extends far beyond that, touching multiple areas of science. In particular, they are subject of research in diverse areas of physics, including high energy physics, nonlinear optics and condensed matter physics \cite{shnir2018topological,abdullaev2014optical,vachaspati2006kinks,kivshar2003optical,rajaraman1982solitons}. Amongst the most known solitons are skyrmions and domain walls in magnetic materials \cite{fert2017magnetic,parkin2008magnetic,koyama2011observation}, vortices in superconductors and fluids \cite{fetter2018theory,kleckner2013creation,auslaender2009mechanics,abrikosov2004nobel} as well as magnetic monopoles, Q-balls, cosmic strings and instantons in high-energy physics \cite{polyakov2018nucleon,schneider2018discrete,csaki2018kaluza,kimball2018searching,
hindmarsh2016new,schaefer2002instanton,vilenkin2000cosmic,kusenko1998supersymmetric,t1974magnetic}. Besides the theoretical applications of solitons, they play an increasingly important role in technology, e.g., in communications \cite{alcon2009long,marin2017microresonator,haus1996solitons}.

Since the solitons are not isolated objects in most physical systems, their interaction with other fields has been subject to intense research in the literature. Boson and Dirac fields interacting with a soliton are known to affect or even create many intriguing phenomena including vacuum polarization and Casimir effect \cite{mohammadi2016finite,gousheh2013casimir}, superconductivity and Bose-Einstein condensation \cite{semenoff2006stretching,burger1999dark}, localization of fermions in the braneworld scenarios \cite{melfo2006fermion}, charge and fermion number fractionalization \cite{jackiw1976solitons} as well as conducting polymers \cite{su1979solitons}. Massless Dirac fermions behave as the quasiparticles in materials such as graphene and topological insulators \cite{neto2009electronic,qi2011topological}.

Exactly solvable models are considered indispensable tools to explore the physics of a system and the symmetries behind it. In this paper, we introduce a parity breaking model with an analytical soliton solution. The potential includes powers up to sixth order in the scalar field $\phi$, where odd powers of $\phi$ exist alongside the even, causing a parity asymmetry.
In \cite{bazeia2018dirac}, the authors considered a massless Dirac field interacting with a skyrmion-like planar defect in a system that does not respect the parity symmetry. They studied the fermion bound spectrum as well as the scattering of fermions from the localized topological structure and found a closed form for the scattering cross-section for small fermion-skyrmion coupling. Parity or inversion symmetry breaking models with topological solutions are of importance in many areas of physics, for example in the context of superconductivity \cite{watanabe2018group,ishizuka2018odd,ruhman2017odd,wang2016topological,wiegmann1990parity}, fractional quantum
Hall effect \cite{read2000paired}, mesoscopic electron transport \cite{deo1995general}, current of abnormal parity \cite{fradkin1986physical},  heavy-ion collisions \cite{abelev2009azimuthal}, nonlinear Schr\"odinger equations \cite{yang2016stability} and hydrodynamics \cite{lucas2014phenomenology}.

In this paper, we consider a parity-breaking model with two minima where the soliton solutions connect the two vacua in a nonsymmetric form. Unlike the kink of $\phi^4$ theory, they have a power-law tail at one side and exponential asymptotics at the other. This behavior can be found most frequently in models where the potential has higher than sextic power in the scalar field \cite{lohe1979soliton}. These types of solitons are especially interesting in the context of the soliton-soliton interactions (see, e.g., \cite{manton2019forces,christov2019long,christov2019kink,belendryasova2019scattering}). Although this is not the main focus of this work, we will comment on it when we find it relevant. The goal here is to find the soliton solutions and stability equation analytically as well as to study the interaction of the soliton with boson and fermion fields. We consider two types of interactions with the boson field and Yukawa interaction for the fermion field. The boson bound and scattering states, as well as the fermion zero mode, are expressed in closed analytical forms. However, the massive fermion bound states and energy spectrum are solved numerically. Most analytical results are expressed in terms of the Lambert W function. In Sec.~2, we introduce the model, find the corresponding topological solution and analyze the small oscillations of the soliton. In Sec.~3, we study the interaction of boson and fermion fields with the soliton of our model. 
Finally, in Sec.~4, we summarize the results of the current work. The appendices provide the details of the calculations.

\section{Model}
We propose the theory described by the following Lagragian
\begin{align}\label{eq:lagrangian}
	\mathcal L = \frac{1}{2}\partial_\mu\phi\,\partial^\mu\phi - V(\phi)\,,
\end{align}
where the potential term is given by
\begin{align}\label{eq:V}
	V(\phi)=\frac{\lambda^2}{2}\left(1-\phi^2\right)^2\left(1-\phi\right)^2\,.
\end{align}
The potential presents two minima, $\phi_0=\pm1$, which allows one to obtain solitonic solutions interpolating between them. 
Although the potential is sixth-order, it has no parity symmetry since odd powers of $\phi$ are also included, unlike the classical $\phi^6$ theory. An equivalent potential could be considered by mapping $\phi\to-\phi$, resulting in an interchange of the roles of the kink and antikink solutions.
The coupling $\lambda$ has mass dimension one, defining a natural mass scale in the system, which we use to rescale all the parameters. Nevertheless, when deemed relevant, we explicit the mass dimension as a function of $\lambda$.

\begin{figure}[htp]
\centering
\begin{tabular}{cc}
	\includegraphics[width=0.45\textwidth]{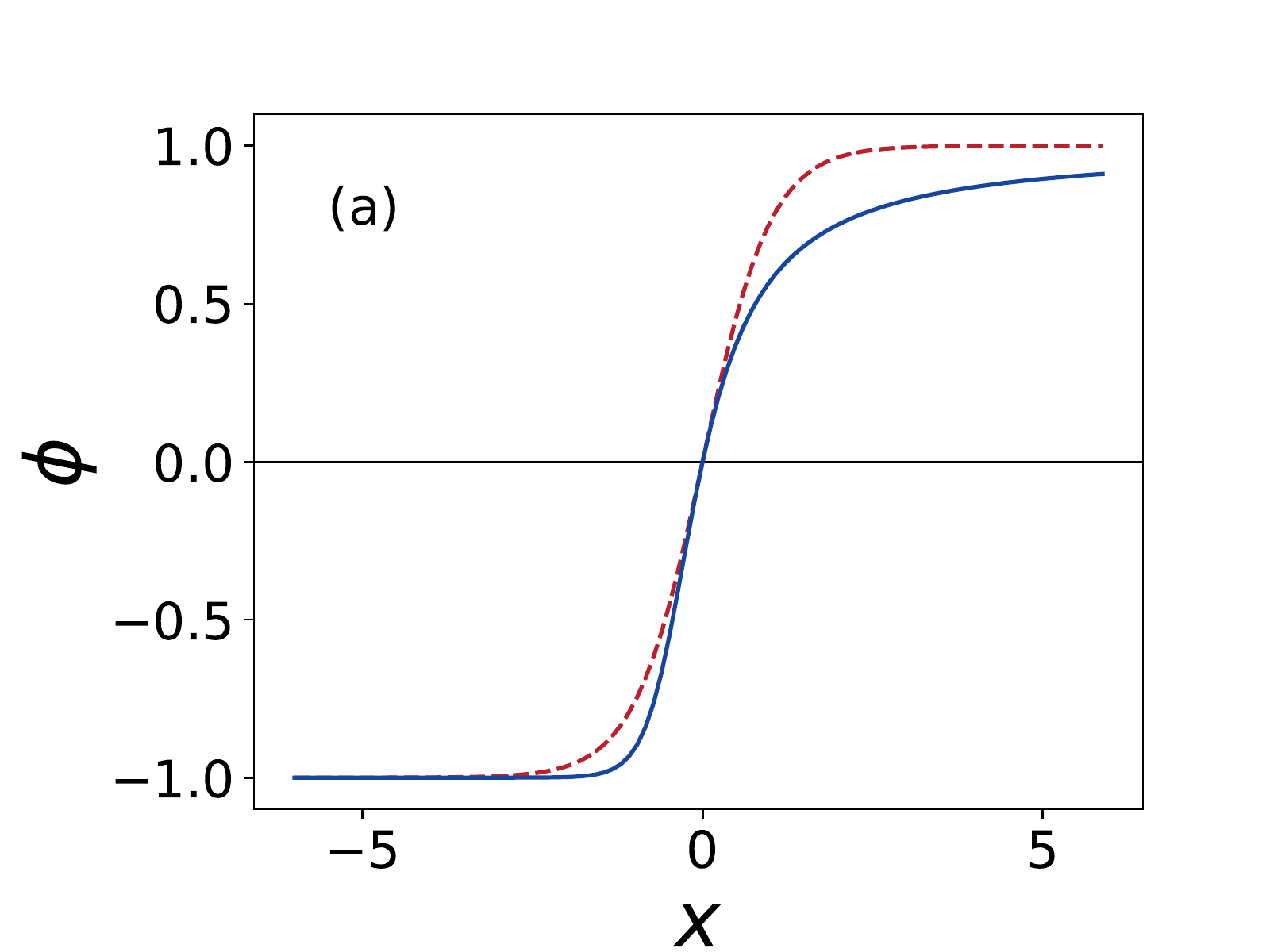} & \includegraphics[width=0.45\textwidth]{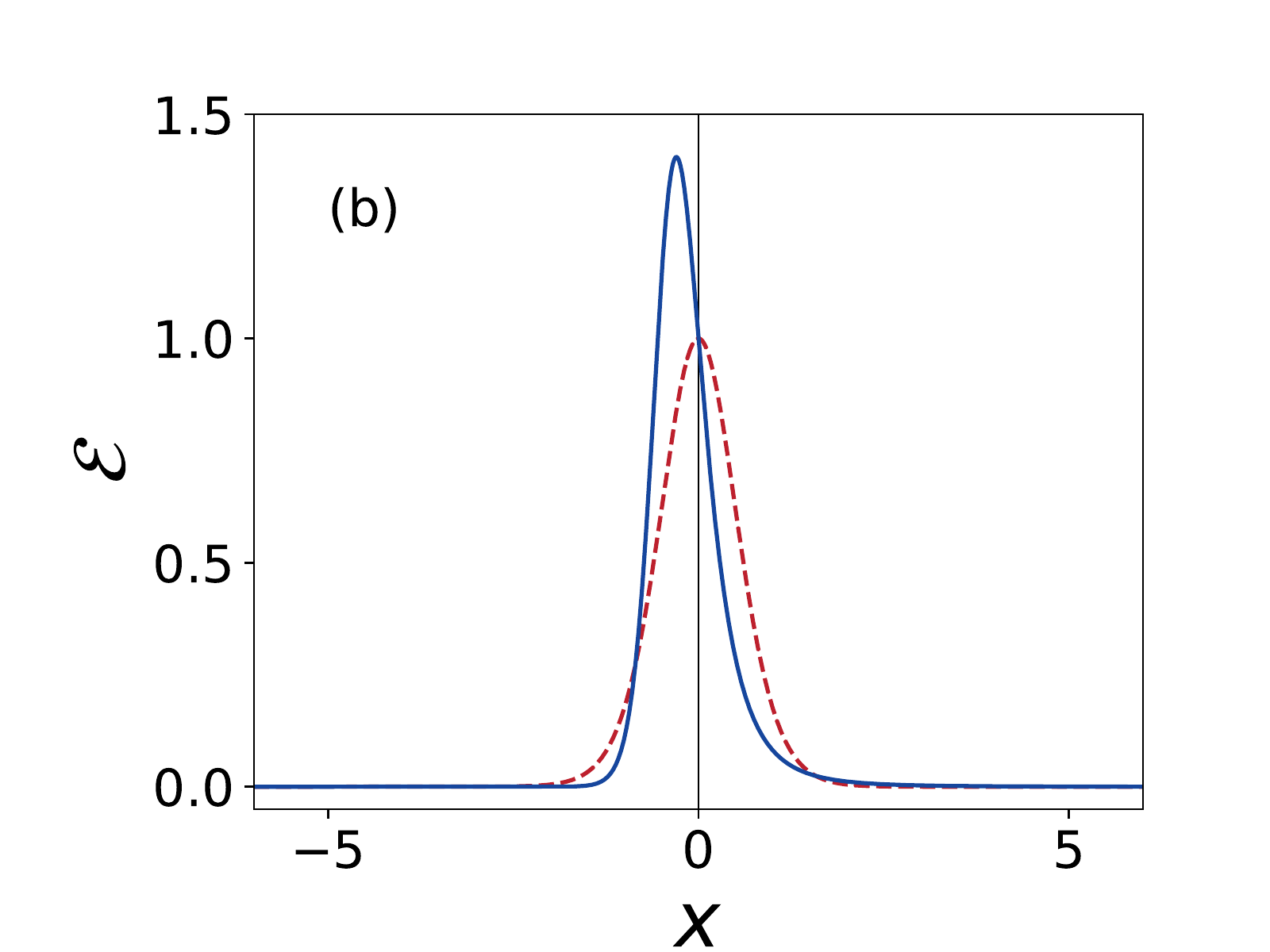}
\end{tabular}
\caption{(a) Soliton profile. (b) Energy density.
The solid line (blue) and the dashed line (red) show the soliton in our model and the kink of $\phi^4$ theory, respectively.}
\label{fig:kinkprofile}
\end{figure}

Although Lagragian (\ref{eq:lagrangian}) yields a second order equation of motion, thanks to the BPS condition one can obtain an equivalent first order equation
\begin{align}
	\partial_x\phi - \left(1-\phi^2\right)\left(1-\phi\right) = 0\,,
\end{align}
where the field $\phi$ is static. 
Integrating the above equation we find 
\begin{align}\label{phix}
	\frac{1}{4} \left[\log\left(\frac{\phi+1}{\phi-1}\right) - \frac{2}{\phi-1}\right] = x + C\,,
\end{align}
where $C$ is the integration constant. We choose the center of the soliton at $\phi(0)=0$, implying $C=\frac{1}{4} (2-i\pi)$. 
This can be solved in terms of Lambert $W$ function\footnote{For the properties of Lambert $W$ function check, e.g., \cite{lambertw1996}.}. The details of this calculation are provided in Appendix~\ref{app:p2}. The solution is
\begin{align}
	\phi_s(x)=1-\frac{2}{1+W\left[e^{1+4x}\right]}.
\end{align}
The corresponding antikink solution can be obtained by mapping $\phi\to -\phi$. Figure~\ref{fig:kinkprofile}a shows the kink profiles for $\phi^4$ and our models. Notice that in the kink profile of our model the parity is explicitly broken.
At large $x$ the behavior of the kink is as follows
\begin{align}
\begin{cases}
   \phi_s(x) \to -1+2e^{1+4x} & \quad x\to -\infty,\\
\phi_s(x) \to 1-1/(2x) & \quad x\to \infty.
\end{cases}
\end{align}
The above asymptotic behavior means that the kink at large $x$ has a long-range power-law fall-off in contrast with the opposite tail, $x \to -\infty$, with exponential asymptotics.

Using the BPS condition, it is straightforward to calculate the energy of the soliton configuration, the so-called classical mass of the soliton,
\begin{align}
	M_{cl} &= \int_{-\infty}^{\infty}{\mathcal E}(x){\mathrm{d}x} \nonumber\\&= \int_{-\infty}^{\infty}\left[ \frac{1}{2}\left(\frac{\mathrm{d}\phi}{\mathrm{d}x}\right)^2 + V(x) \right]\mathrm{d}x = \int_{-1}^{1}(1-\phi^2)^2(1-\phi)^2 \ \mathrm{d}\phi = \frac{4}{3}\,(\lambda)\,,
\end{align}
where the energy density ${\mathcal E}(x)$ is shown in Fig.~\ref{fig:kinkprofile}b for our model and $\phi^4$ kink.
Interestingly, despite the difference in the energy density of the two models, the resulting mass is the same.

Having the profile of the soliton, it is relevant to analyze the small fluctuations of the boson field described by the linear stability equation
\begin{align}
\left[-\partial_x^2+U[\phi(x)]\Bigr|_{\phi_s(x)}\right] \eta_n (x)=\omega_n^2 \eta_n (x)\, ,
\end{align}
with the stability potential
\begin{align}
U[\phi(x)]\Bigr|_{\phi_s(x)}&=\left.\frac{d^2 V}{d \phi^2}\right|_{\phi_s(x)}\nonumber\\&= \frac{16\left(1-8 W\left[e^{1+4x}\right]+6\left(W\left[e^{1+4 x}\right]\right)^2\right)}{\left(1+W\left[e^{1+4x}\right]\right)^4},
\end{align}
where $\eta_n$'s are the normal modes of the fluctuations around the static solution.
Due to the translational symmetry of the system there exists a zero mode, $\omega_0=0$. It is possible to show that it is as follows
\begin{align}
\eta_0=\partial_x \phi=\frac{8 W\left[e^{1+4x}\right]}{\left(W\left[e^{1+4x}\right]+1\right)^3}.
\end{align}

\begin{figure}[htp]
\centering
\begin{tabular}{cc}
	\includegraphics[width=0.45\textwidth]{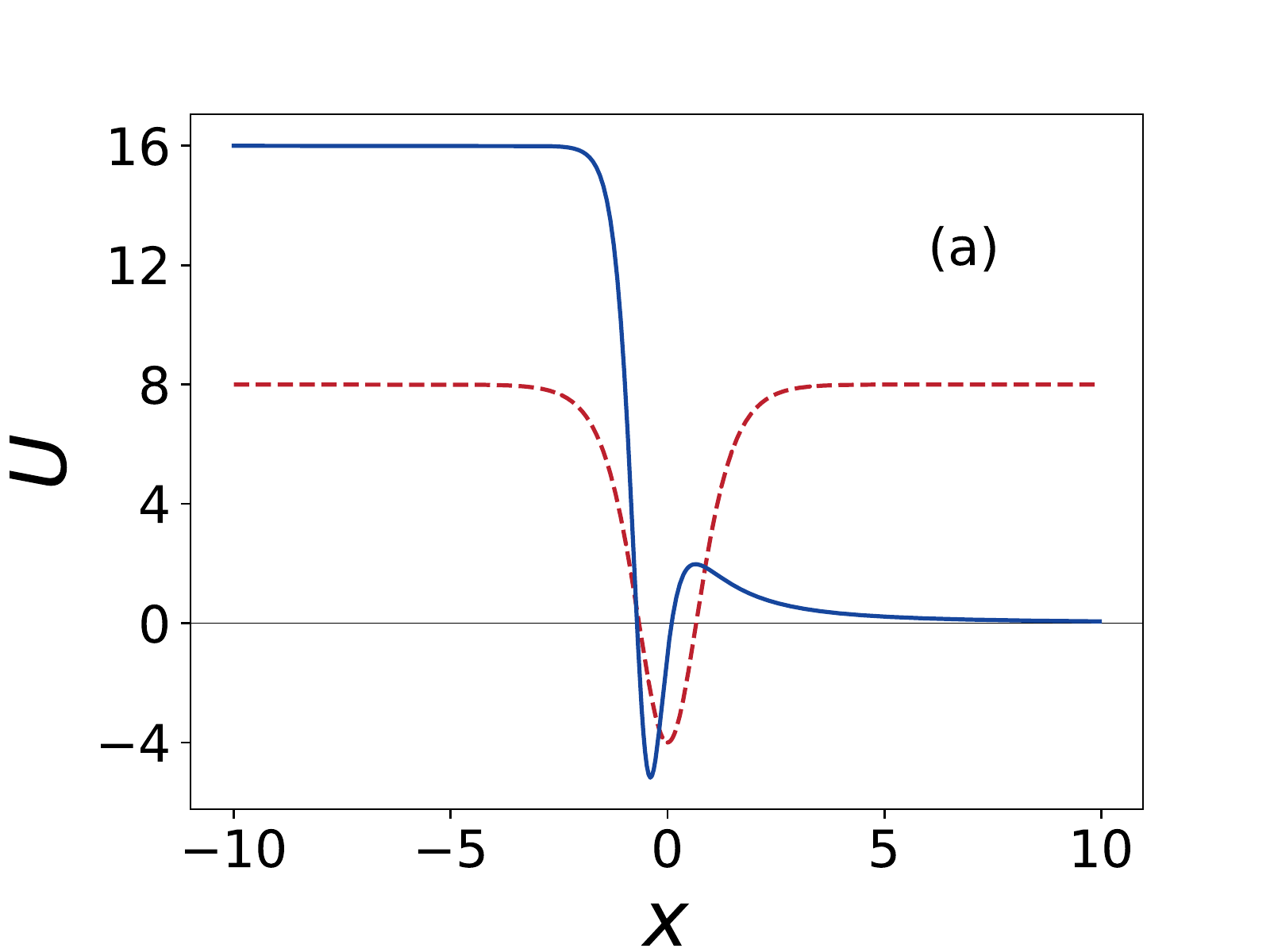} & \includegraphics[width=0.45\textwidth]{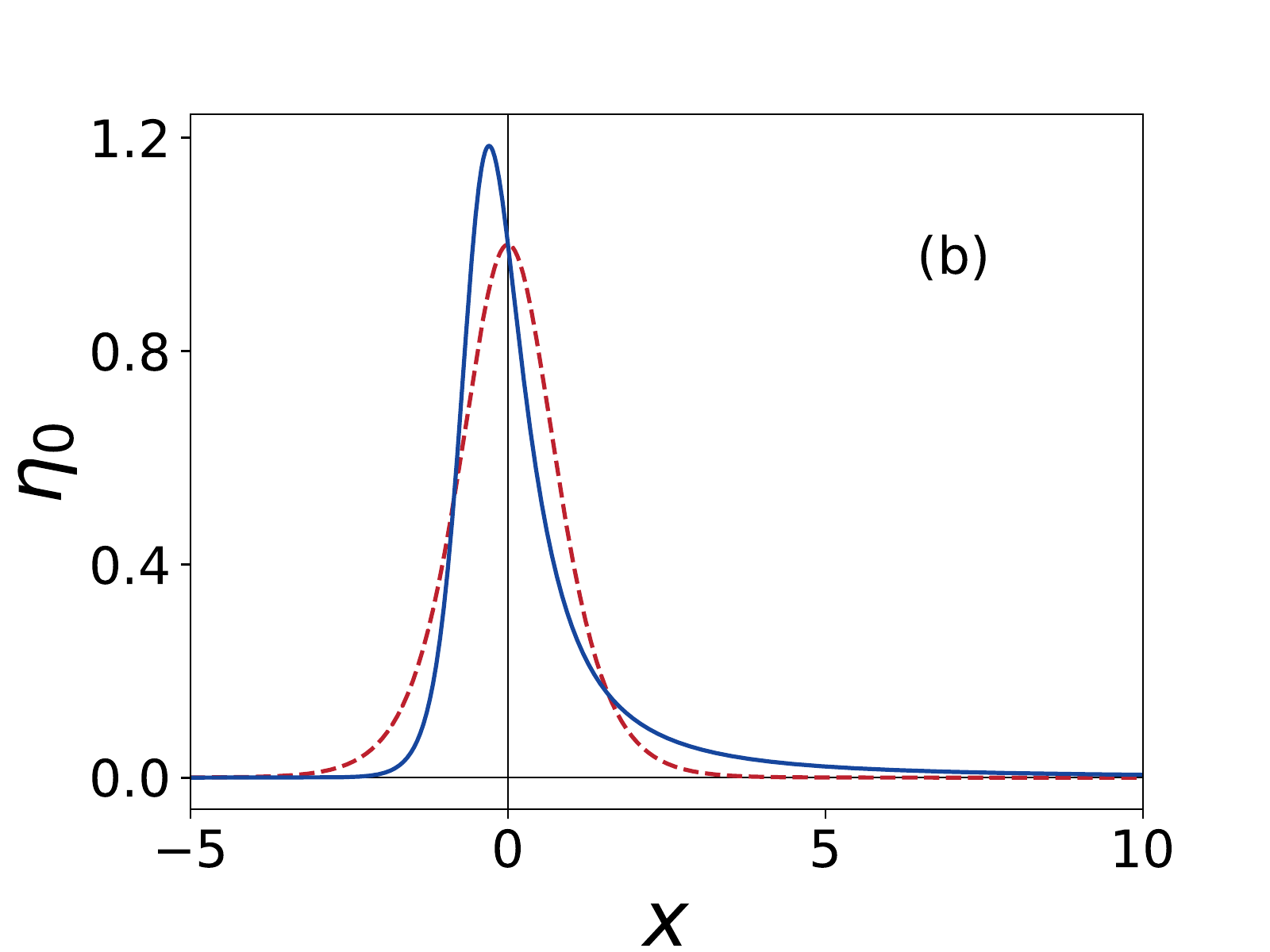}
\end{tabular}
\caption{(a) Stability potential. (b) Zero mode. In both cases, the solid line (blue) and the dashed line (red) show the graphs for our model and the kink of $\phi^4$ theory, respectively.}
\label{fig:stability-potential}
\end{figure}

Knowing that
\begin{align}
\left.\frac{d^2 V}{d \phi^2}\right|_{\phi=1}=0 \,, \quad\quad\quad\quad
\left.\frac{d^2 V}{d \phi^2}\right|_{\phi=-1}=16\,(\lambda^2)\,,
\end{align}
and requiring $\omega_n^2$ to be non-negative, implies that the zero mode is the only discrete mode, i.e., there are no bound oscilation modes of the soliton apart from translation. Figure~\ref{fig:stability-potential} shows the stability potential $U(x)$ (panel (a)) as well as the zero mode $\eta_0$ (panel (b)). As one can see, the potential presents different limits at $x\to\pm\infty$. Due to this fact, only waves whose energy exceeds $U(-\infty)$ are permitted when travelling from the left. In contrast, incoming waves from the right are allowed for lower energies starting from 0, the value of $U(+\infty)$. In this case, oscillations with an energy smaller than $U(-\infty)$ are totally reflected by the potential barrier. 

Until now, we have been concerned with the soliton solutions in isolation. In what follows, we consider the interaction of the soliton of our model with other fields, including boson and fermion fields. We analyze two different types of couplings responsible for the soliton-boson interaction, and a Yukawa coupling between the soliton and the Dirac field. In all three cases, we consider the soliton a background field.

\section{Interaction with a scalar field}
\subsection{Model I}\label{ssec:model1}
First, let us consider the interaction of a real massive scalar field $\chi$ with the soliton of our model in the following form
\begin{align}
	\mathcal L = \frac{1}{2}\partial_\mu\phi\,\partial^\mu\phi - V(\phi) + \frac{1}{2}\partial_\mu\chi\,\partial^\mu\chi + \frac{1}{2}m^2\chi^2 - g\,\phi\,\chi\,,
\end{align}
where $m$ is the mass of the field $\chi$ and $g$ is the scalar-soliton coupling constant.
This interaction yields a non-homogeneous Klein-Gordon equation  
\begin{align}
	\left(\square-m^2\right)\chi=-g\,\phi\,.
\end{align}
Separating the time dependence as $\chi = \chi_se^{-iEt}$, we find the equation
\begin{align} \label{diffeqmodelI}
	\left(\partial_x^2+k^2\right)\chi_s=g\,\phi\,,
\end{align}
where we define $k^2 \equiv E^2-m^2$. First consider the bound states, for which we have $k^2<0$. The solution to the equation of motion, eq.~(\ref{diffeqmodelI}), is
\begin{align}
	\chi_s(x) &= A e^{\kappa x}+B e^{-\kappa x} + \frac{g}{\kappa}\int^x \sinh\left[\kappa(x-y)\right]\phi\left(y\right) \, \textrm{d}y\,,\end{align}
introducing $k^2\equiv -\kappa^2$. The first two terms come from the solution of the homogeneous equation and the last one is a particular solution. Focusing only on the integral term in the above solution and performing the change of variables $u=x-y$ results in
\begin{align}
	& -\frac{g}{\kappa}\int \sinh\left(\kappa u\right)\left(1-\frac{2}{W\left[e^{1+4(x-u)}\right]+1}\right) \, \textrm{d}u
    = - \frac{g}{\kappa^2}
	+\frac{2g}{\kappa}\int \frac{\sinh\left(\kappa u\right)}{1+W\left[e^{1+4 (x-u)}\right]} \, \textrm{d}u\,.
\end{align}
Two more changes of variables, $v=e^{1+4(x-u)}$ followed by $w=W[v]$, allow us to rewrite the integral in a form that can be directly solved

\begin{align}
	& - \frac{g}{\kappa^2}
	-\frac{g}{2\kappa}\int \sinh\left[-\frac{\kappa}{4}\left(\ln(w) + w -1-4x\right)\right]\frac{1}{w} \, \textrm{d}w\nonumber\\
	=&- \frac{g}{\kappa^2}
	-\frac{g}{4\kappa}\left[\left(-\frac{\kappa}{4}\right)^{-\kappa/4} e^{-\frac{\kappa}{4}(1+4x)}\,\Gamma \left(\frac{\kappa}{4},-\frac{\kappa}{4} W\left[e^{1+4x}\right]\right)\right.\nonumber\\-&\left. \left(\frac{\kappa}{4}\right)^{\kappa/4} e^{\frac{\kappa}{4} (1+4x)}\,\Gamma\left(-\frac{\kappa}{4},\frac{\kappa}{4} W\left[e^{1+4x}\right]\right)\right]\,.
\end{align}
Therefore, the general solution takes the form
\begin{align}\label{Model1Bound}
	\chi_s(x) &= A e^{\kappa x}+B e^{-\kappa x} - \frac{g}{\kappa^2}
	-\frac{g}{4\kappa}\left[\left(-\frac{\kappa}{4}\right)^{-\kappa/4} e^{-\kappa x}\, e^{-\kappa/4}\ \Gamma \left(\frac{\kappa}{4},-\frac{\kappa}{4}\, W\left[e^{1+4x}\right]\right)\right.\nonumber\\-&\left.\left(\frac{\kappa}{4}\right)^{\kappa/4} e^{\kappa x}\, e^{\kappa/4}\ \Gamma\left(-\frac{\kappa}{4},\frac{\kappa}{4}\, W\left[e^{1+4x}\right]\right)\right]\,.
\end{align}
To obtain a real $\chi_s$ one has to impose the restriction $\kappa=4n$ where $n$ is an integer number. This means that $E^2=m^2-16n^2$ and also $n<m/4(\lambda)$ following the fact that $E^2$ is non-negative. Looking at the limit of $\chi_s$ when $x\to+\infty$ it is easy to see that, for the solution to be finite, $A$ should be null. At this limit, the last term in the above expression vanishes and therefore the solution converges to $-g/\kappa^2$. It remains to determine the value of $B$ which can be found by requiring the solution to be finite when $x\to-\infty$. Doing so, it can be shown that
\begin{align}
B=\frac{g}{16n} \left(-n\right)^{-n}\, e^{-n}\ \Gamma \left(n\right)  .
\end{align}The detailed calculations are provided in Appendix \ref{app:model1}.
Figure~\ref{figModel1Bound} shows the bound states for three values of $n$, 1, 2 and 3. As it can be seen, at the limits $x\to\pm\infty$ the solution converges to $\mp\frac{g}{\kappa^2} = \mp\frac{g}{16n^2}$. This result is expected considering eq.~(\ref{diffeqmodelI}) when $\phi(x\to\pm \infty)=\pm 1$.
\begin{figure}[htp]
\centering
\includegraphics[width=0.6\textwidth]{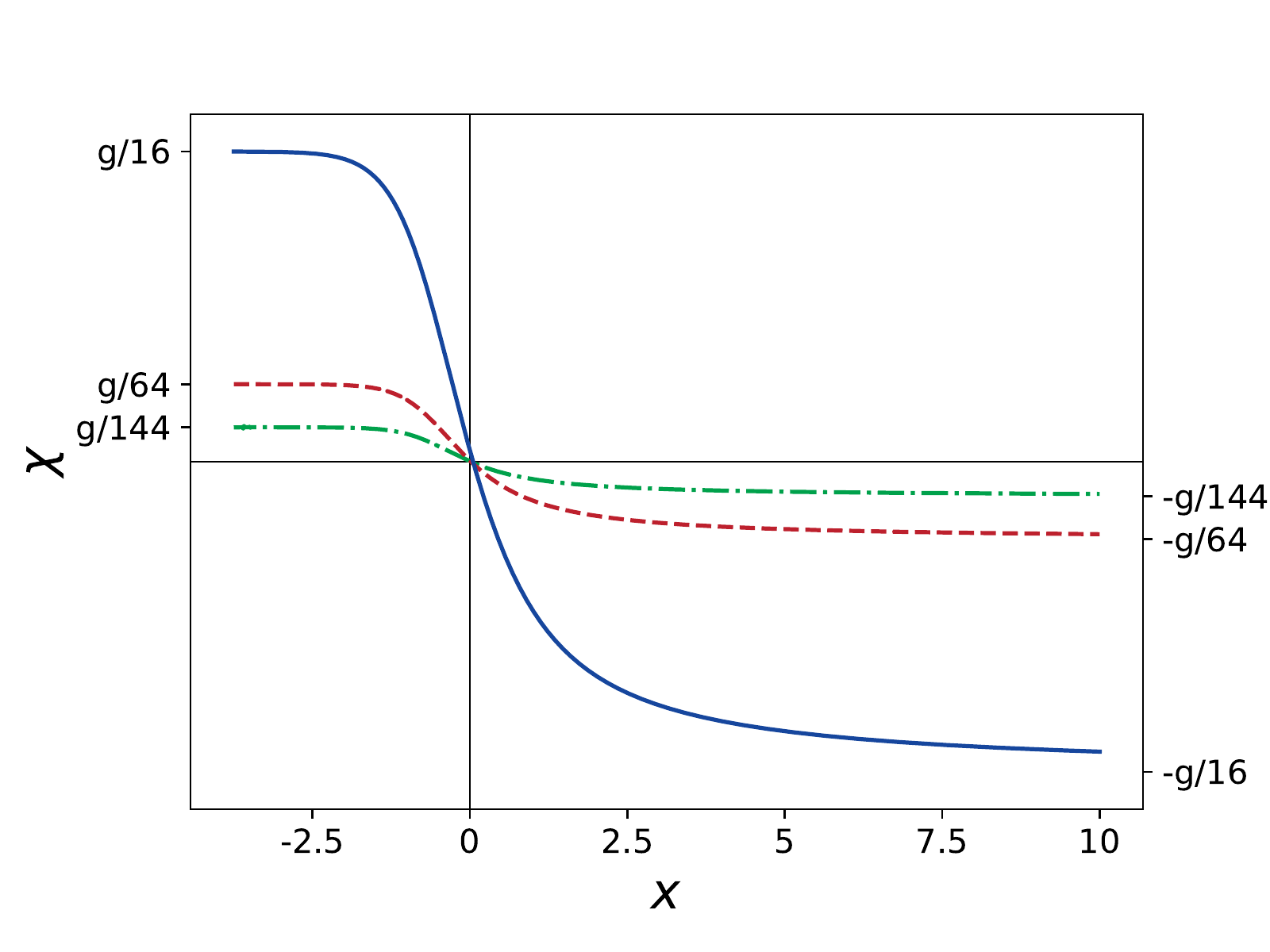}
\caption{Bound states for three values $n=1,2,3$.}
\label{figModel1Bound}
\end{figure}
Notice that, interstingly, bound states are also solitons which means that the original soliton can trap another boson field in the form of a soliton configuration.

Now, let us look at the case $k^2>0$ which corresponds to the scattering states. 
The general solution for this equation is in the form
\begin{align} \label{chimodelI}
\chi_s(x)= A e^{ikx} + B e^{-ikx} + \frac{g}{k}\int^x \sin[k(x-y)]\,\phi(y)\,\textrm{d}y\,,
\end{align}
where again the first two terms come from the solution of the homogeneous equation and the last one is a particular solution.
Following the same series of change of variables the integral in the above expression changes to
\begin{align}
	&-\frac{g}{2k}\int \sin\left[-\frac{k}{4}\left(\ln(w) + w -1-4x\right)\right]\,\frac{1}{w}\textrm{d}w\nonumber\\
	=&\frac{-i g}{4k} \left[\Gamma(-i k/4,i kw/4) (i k/4 )^{i k/4}\, e^{i(1+4x)k/4} - c.c.\right]\,,
\end{align}
where $c.c.$ stands for the complex conjugate. After some simplifications, the solution (\ref{chimodelI}) takes the form
\begin{align} \label{solmodelI}
	\chi_s(x)&=A e^{ikx} + B e^{-ikx} + \frac{g}{k^2}+\nonumber\\&+\frac{g}{2k} \, \mathrm{Im}\left[\Gamma(-i k/4,i k\,W[e^{1+4x}]/4)\, e^{i[1+4x+\ln(k/4)+i\pi/2]k/4}\right]\,.
\end{align}
To verify the result, one can look at the limits $x\to\pm\infty$. At the limit $x\to+\infty$, the last term in the above solution tends to zero and we recover the expected result using eq.~(\ref{diffeqmodelI}) when $\phi\to1$ 
\begin{align}
	\chi_s(x\to+\infty)&=A e^{ikx} + B e^{-ikx} + \frac{g}{k^2}\,.
\end{align}
The same goes for the limit $x\to-\infty$ where the eq. (\ref{solmodelI}) tends to
\begin{align}
	\chi_s(x\to-\infty)&=A e^{ikx} + B e^{-ikx} - \frac{g}{k^2}\nonumber\\&+\frac{g}{2k} \,\textrm{Im}\left[\left(\frac{k}{4}\right)^{i k/4} e^{(\pi +i) k/4}\, \Gamma \left(-i \frac{k}{4}\right) e^{i k x}\right]\,.
\end{align}
The last term can be removed through a redefinition of the coefficients $A$ and $B$, which gives the expected result using eq.~(\ref{diffeqmodelI}) when $\phi\to -1$.

\subsection{Model II}
\begin{figure}[htp]
\centering
\includegraphics[width=0.45
\textwidth]{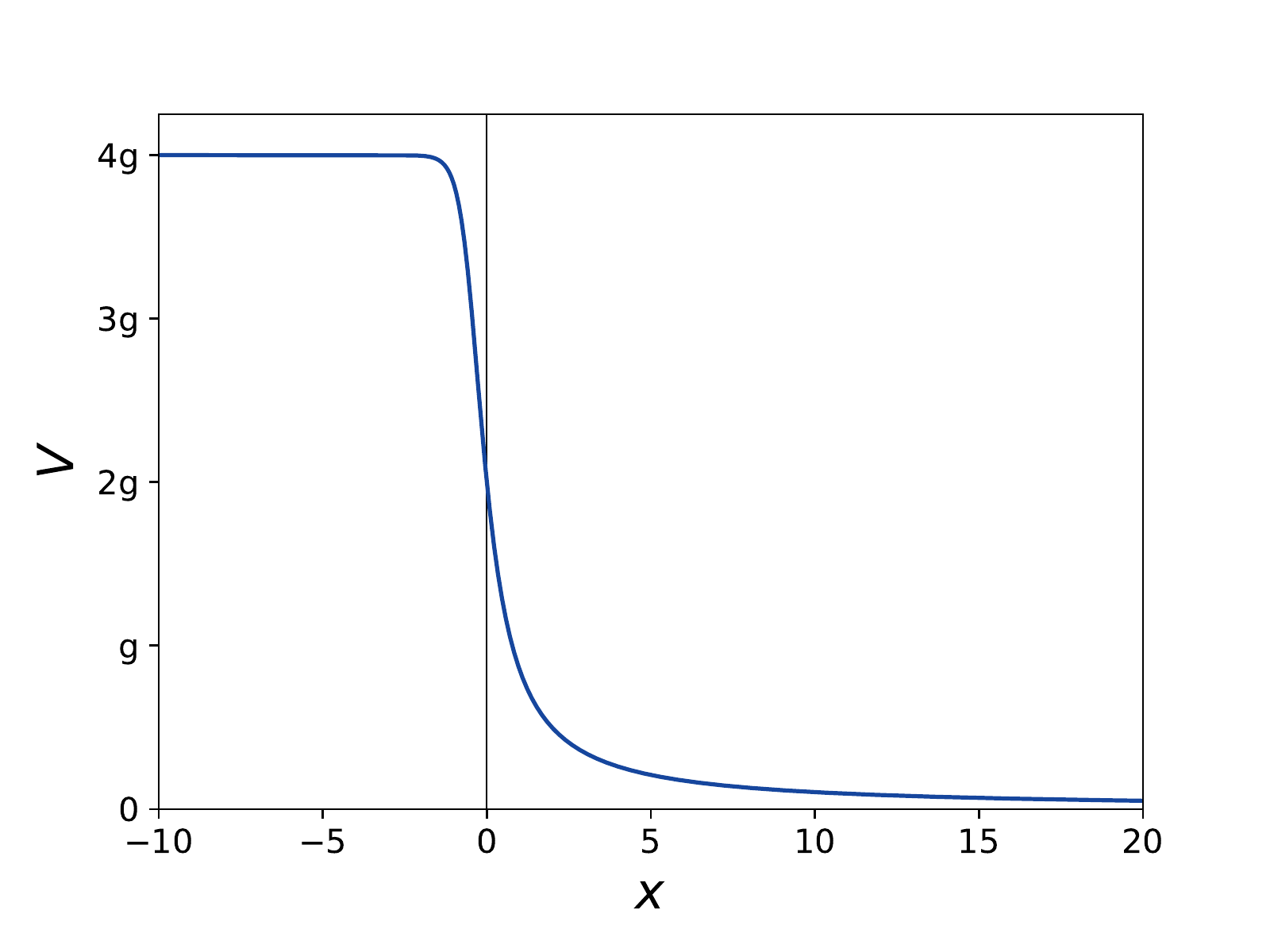}
\caption{Potential energy of the Schr\"odinger-like equation.}
\label{figModel2Potential}
\end{figure}

\begin{figure}[htp]
\centering
\begin{tabular}{cc}
	\includegraphics[width=0.45\textwidth]{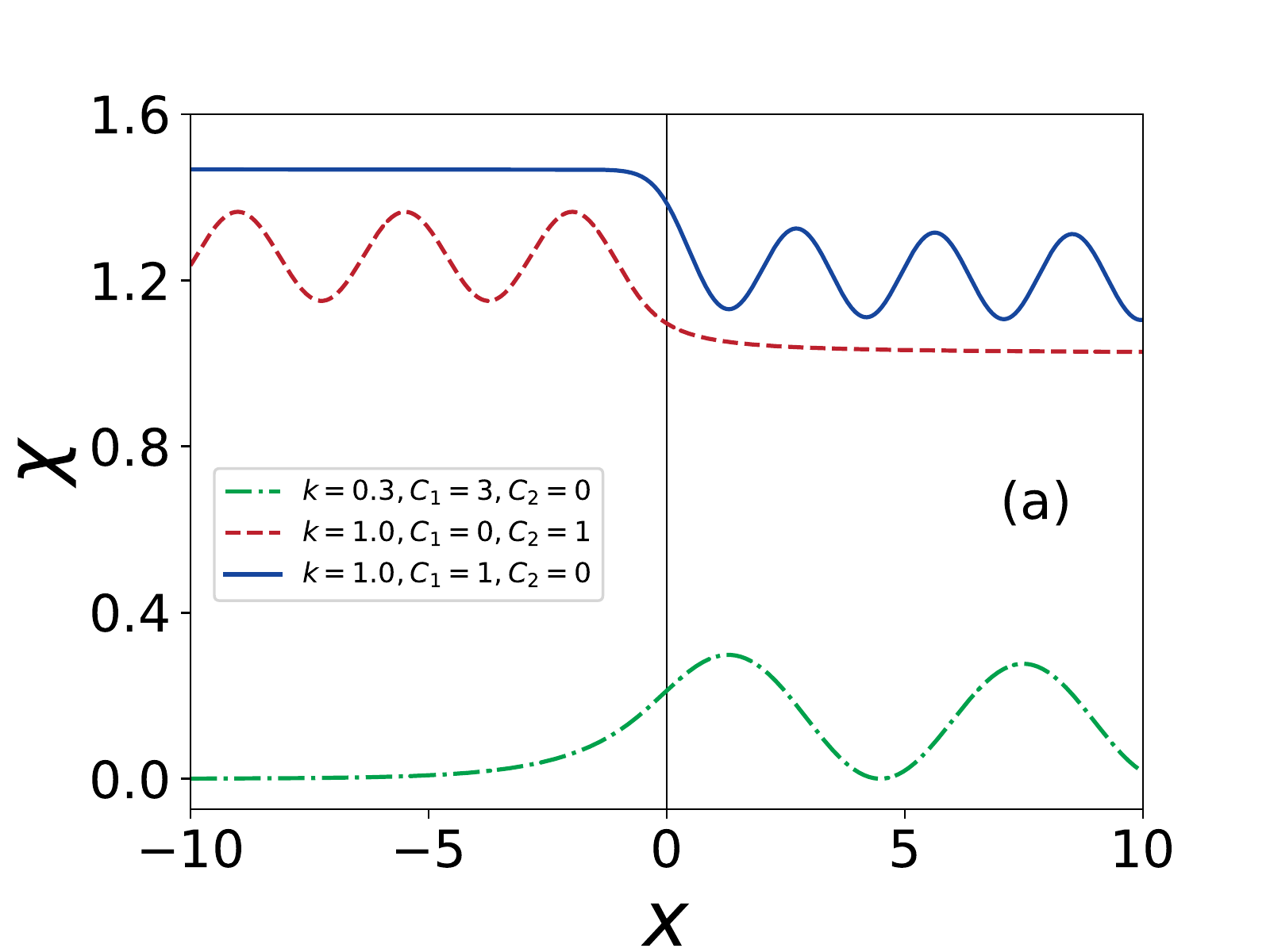} & \includegraphics[width=0.45\textwidth]{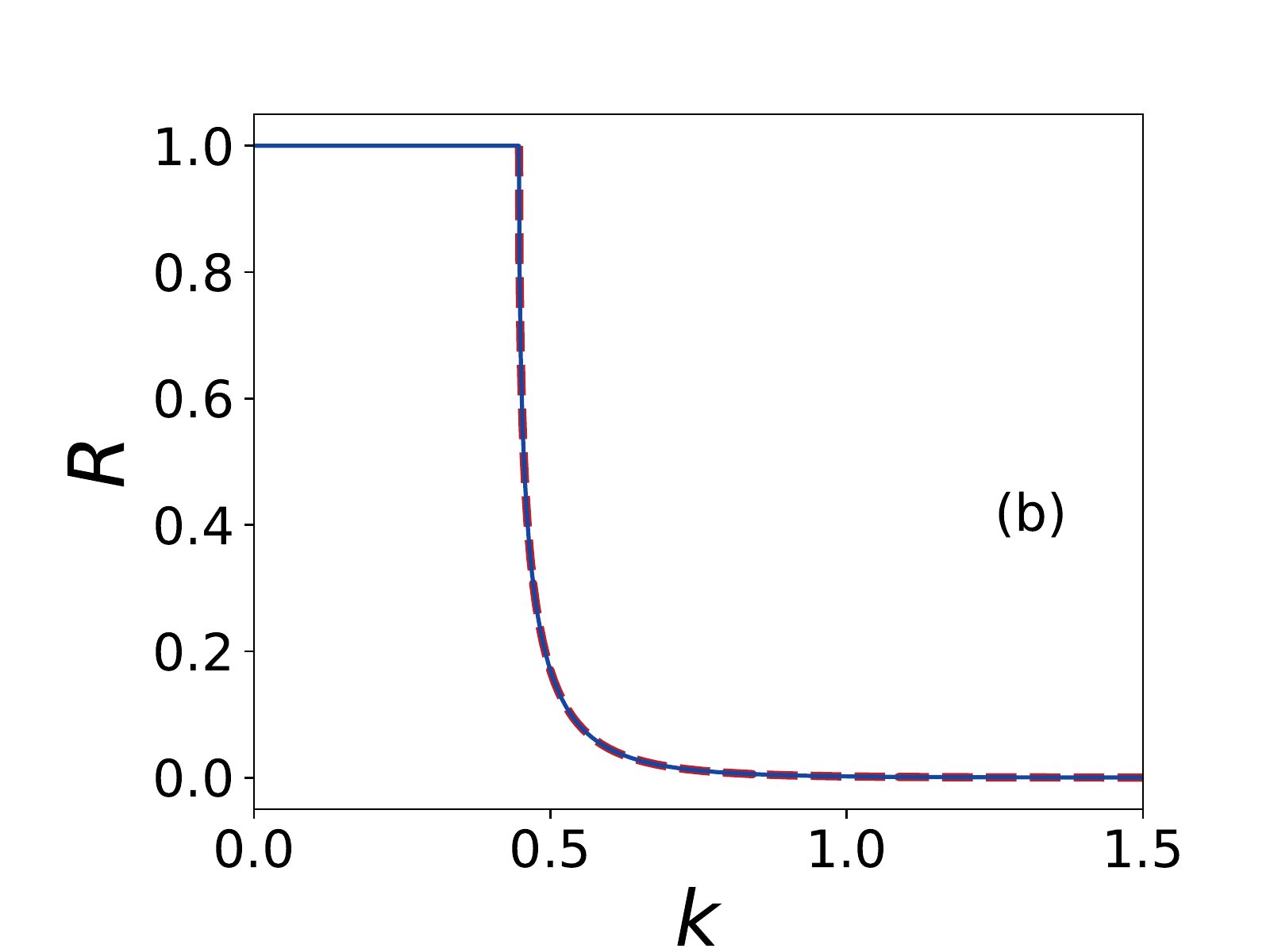}
\end{tabular}
\caption{(a) Boson field continuum states. (b) Reflection coefficient. In both cases, the solid line (blue) and the dashed line (red) show the graphs for scattering from the right and scattering from the left considering $g=0.1$, respectively. The dot-dashed curve (green) shows the scattering from the right for the case where the energy is below the threshold required to surpass the barrier.}
\label{fig:scat-model2}
\end{figure}

Now we introduce a different type of coupling between the soliton field $\phi$ and the scalar field $\chi$. Consider the following Lagragian 
\begin{align}
	\mathcal L = \frac{1}{2}\partial_\mu\phi\,\partial^\mu\phi - V(\phi) + \frac{1}{2}\partial_\mu\chi\,\partial^\mu\chi + \frac{1}{2}m^2\chi^2 + g\,\phi\,\chi^2\,,
\end{align}
where the coupling between the fields is analogous to a Yukawa interaction. This interaction yields the equation of motion 
\begin{align}
	\left(\square-m^2\right)\chi-2g\,\phi\chi=0\,.
\end{align}
Considering $\chi = \chi_se^{-iEt}$ and rearranging the terms we arrive at 
\begin{align}
	\left(-\partial_x^2 - 2g\,\phi\right)\chi_s = k^2 \chi_s\,.
\end{align}
Replacing the solitonic solution of our model in the above equation leads to
\begin{align}
	\left(-\partial_x^2 + \frac{4g}{1+W[e^{1+4x}]}\right)\chi_s = (k^2+2g) \chi_s\,,
\end{align}
which has the formal structure of the Schr\"odinger equation with energy equal to $(k^2+2g)$. Figure~\ref{figModel2Potential} shows the form of the potential term in the above Schr\"odinger-like equation. In \cite{ishkhanyan2016lambert}, the author solved a similar equation. To map our system to the quantum mechanical system solved in the aforementioned paper, we need first to consider the change of variables $1+4x \to -y$ which results in
\begin{align}
	\left(-\partial_y^2 + \frac{g/4}{1+W[e^{-y}]}\right)\chi_s = \frac{1}{16}(k^2+2g) \chi_s
\end{align}
Now, the map between their system and ours is given by $2m/\hbar^2 \to 1$, $V_0\to g/4$, $E\to (k^2+2g)/16$ and $\sigma\to1$. As a result, the solution to our system is in the following form
\begin{align}
	\chi_s=z^{i\delta^-/2}e^{-i\delta^+z/2}\left(\frac{\mathrm{d}u(z)}{\mathrm{d}z}-i\,\frac{\left(\delta^{+}+\delta^-\right)}{2} \, u(z)\right)
\end{align}
with $z=W\left[e^{1+4x}\right]$, $\delta^\pm=\frac{1}{2}\sqrt{k^2\pm2g}$, $a=\left(\delta^{+}+\delta^-\right)^2/(4\delta^+)$,
\begin{align}
	u = C_1 \, (i\delta^+z)^{1-i\delta^-}\,_1F_1(1+i(a-\delta^-);2-i\delta^-;i\delta^+z)+C_2 \, U(ia;i\delta^-;i\delta^+z)\nonumber,
\end{align}
where $C_1$ and $C_2$ are constants and $_1F_1$ and $U$ are the Kummer and Tricomi confluent hypergeometric functions, respectively. The system does not have any bound state, which is easy to recognize from the form of the potential (see Fig.~\ref{figModel2Potential}). The scattering states from the right and the left are shown in Fig.~\ref{fig:scat-model2}a. Besides that, in the same figure, one can see the scattering from the right where the energy is beneath the threshold required to surpass the barrier. In this case, the wave is totally reflected. Figure~\ref{fig:scat-model2}b shows the reflection coefficient as a function of the momentum for the waves from the right and left. In the case of incoming waves from the right, the reflection coefficient is 1 for momenta associated with energies below the barrier, as expected. Also, for waves coming from both directions, the reflection coefficient drops to zero at high energies since the wave does not see the barrier.

\section{Interaction with a fermion field}
\begin{figure}[htp]
\centering
\begin{tabular}{cc}
\includegraphics[width=0.45\textwidth]{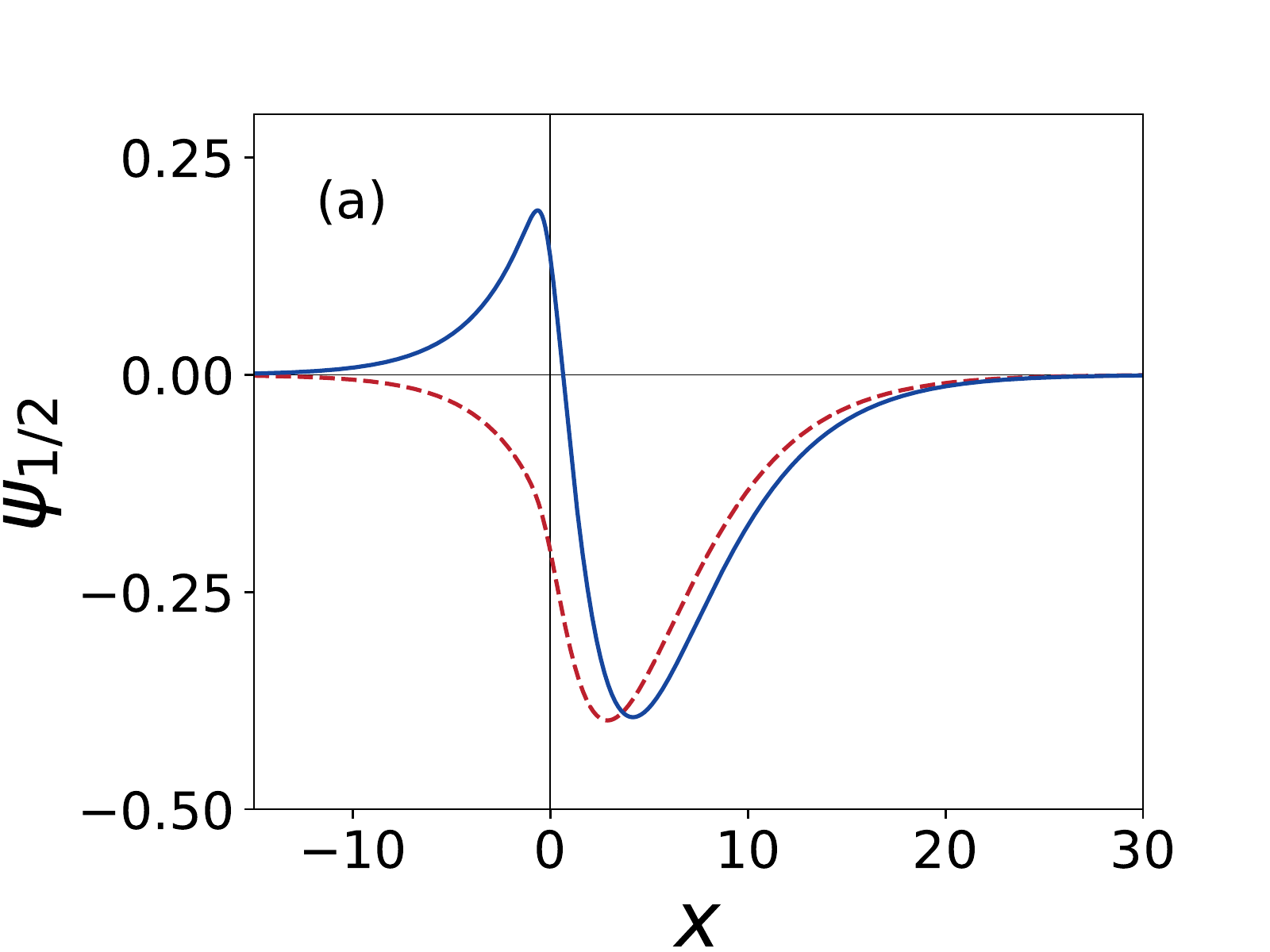}&
\includegraphics[width=0.45\textwidth]{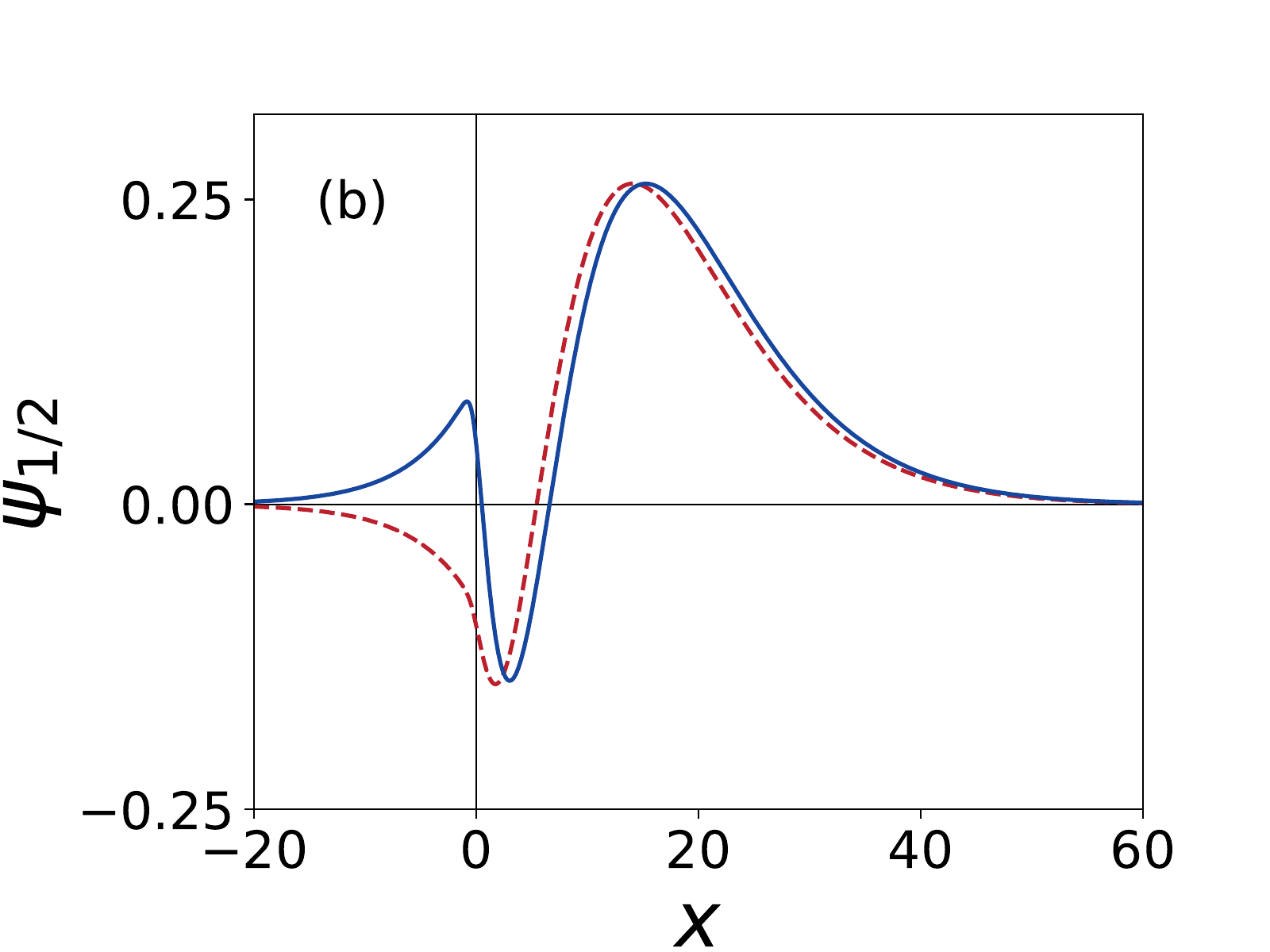}
\end{tabular}
\caption{(a) First fermionic bound state for $g=0.9$. (b) Second fermionic bound state for $g=0.9$. The solid (blue) and dashed (red) curves show the upper and lower components, $\psi_1$ and $\psi_2$, respectively.}
\label{fig:fermionbound}
\end{figure}

\begin{figure}[htp]
\centering
\begin{tabular}{cc}
\includegraphics[width=0.45\textwidth]{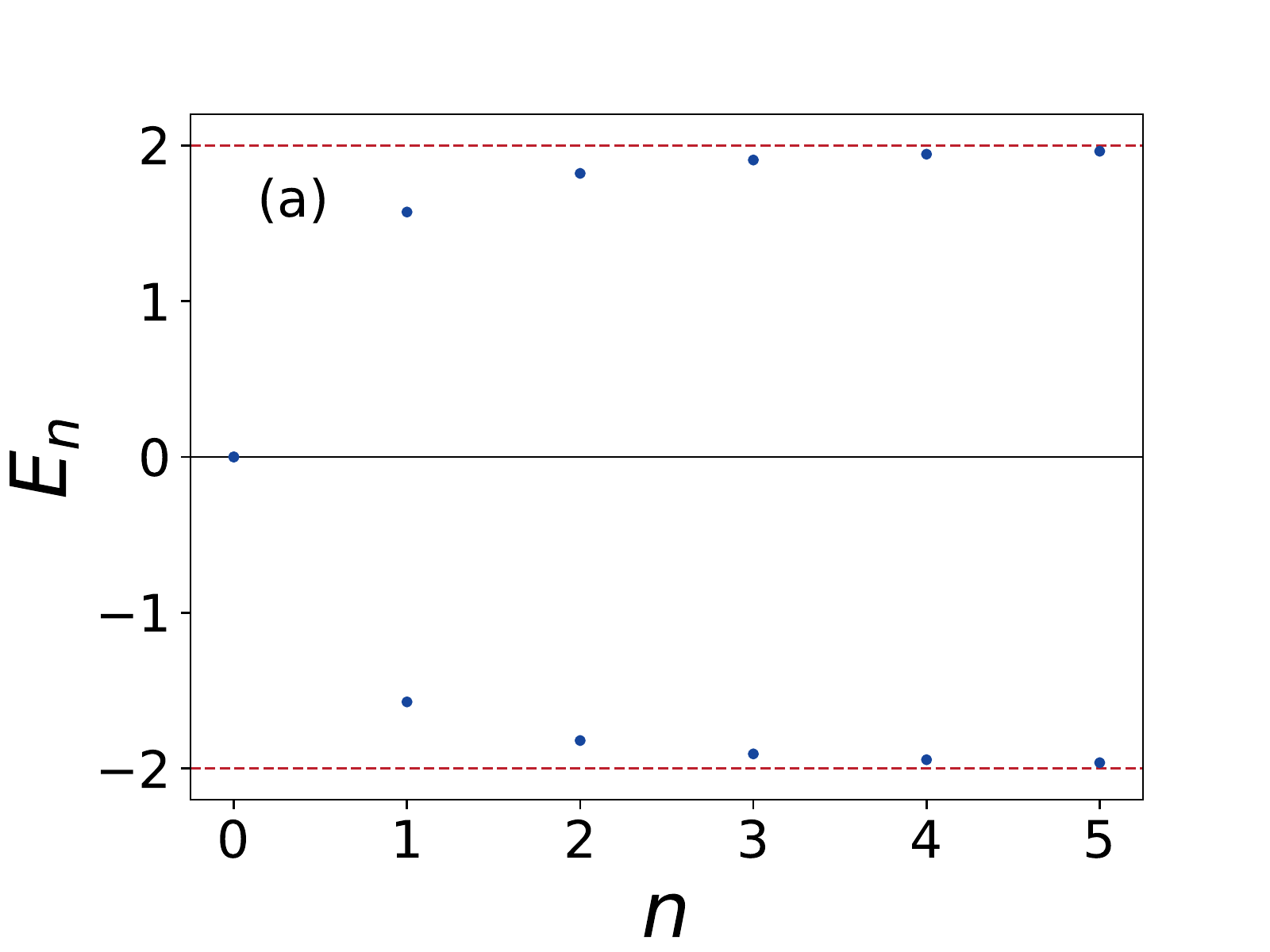} &
\includegraphics[width=0.45\textwidth]{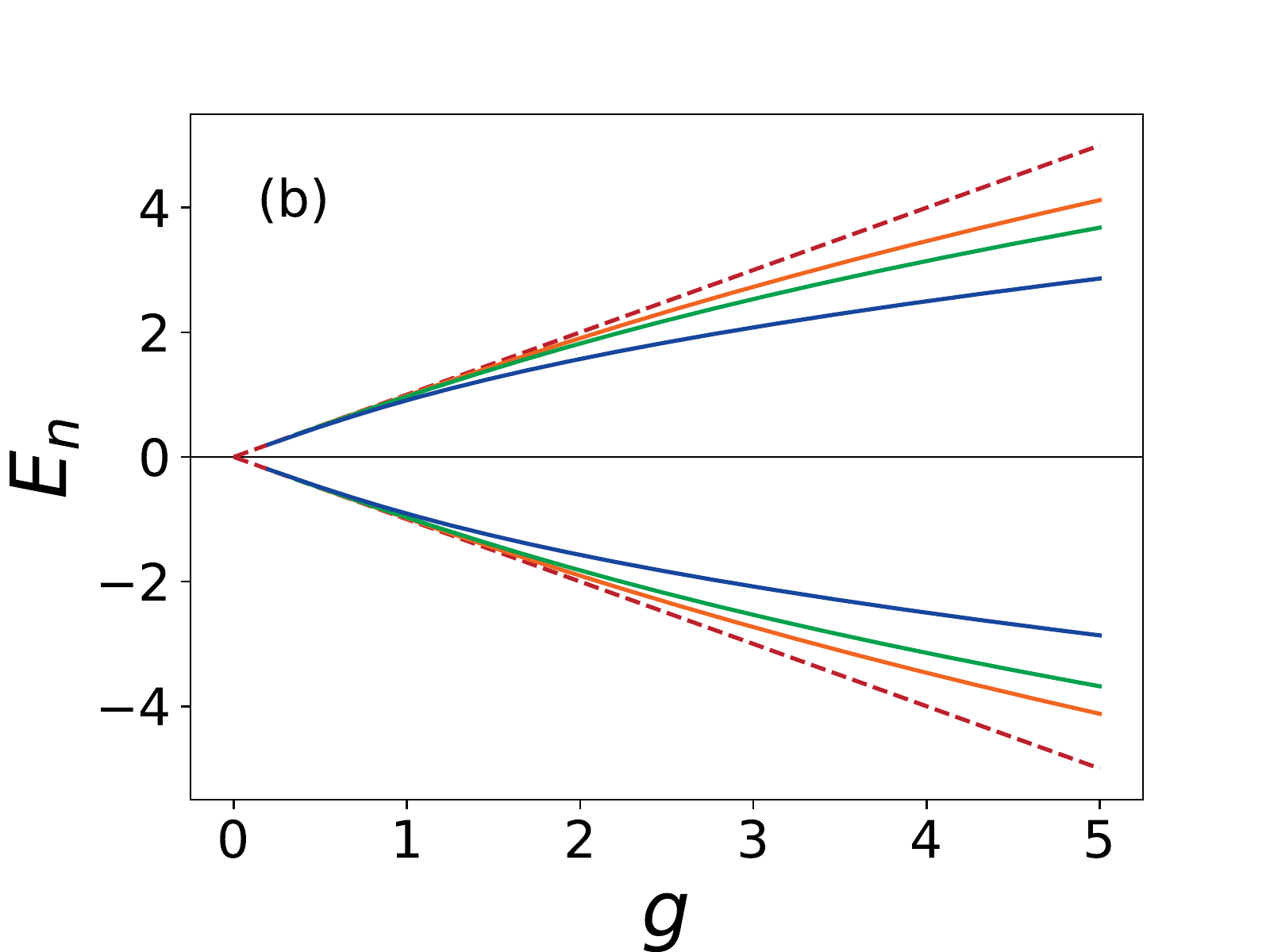}
\end{tabular}
\caption{(a) Fermionic bound energy spectrum for $g=2$. The dashed lines (red) show the threshold energies. (b) Fermionic bound energy spectrum for the first three bound states as a function of the coupling $g$. The dashed lines (red) show the threshold energies.}
\label{fig:fermionspectrum}
\end{figure}

\begin{figure}[htp]
\centering
\begin{tabular}{cc}
\includegraphics[width=0.45\textwidth]{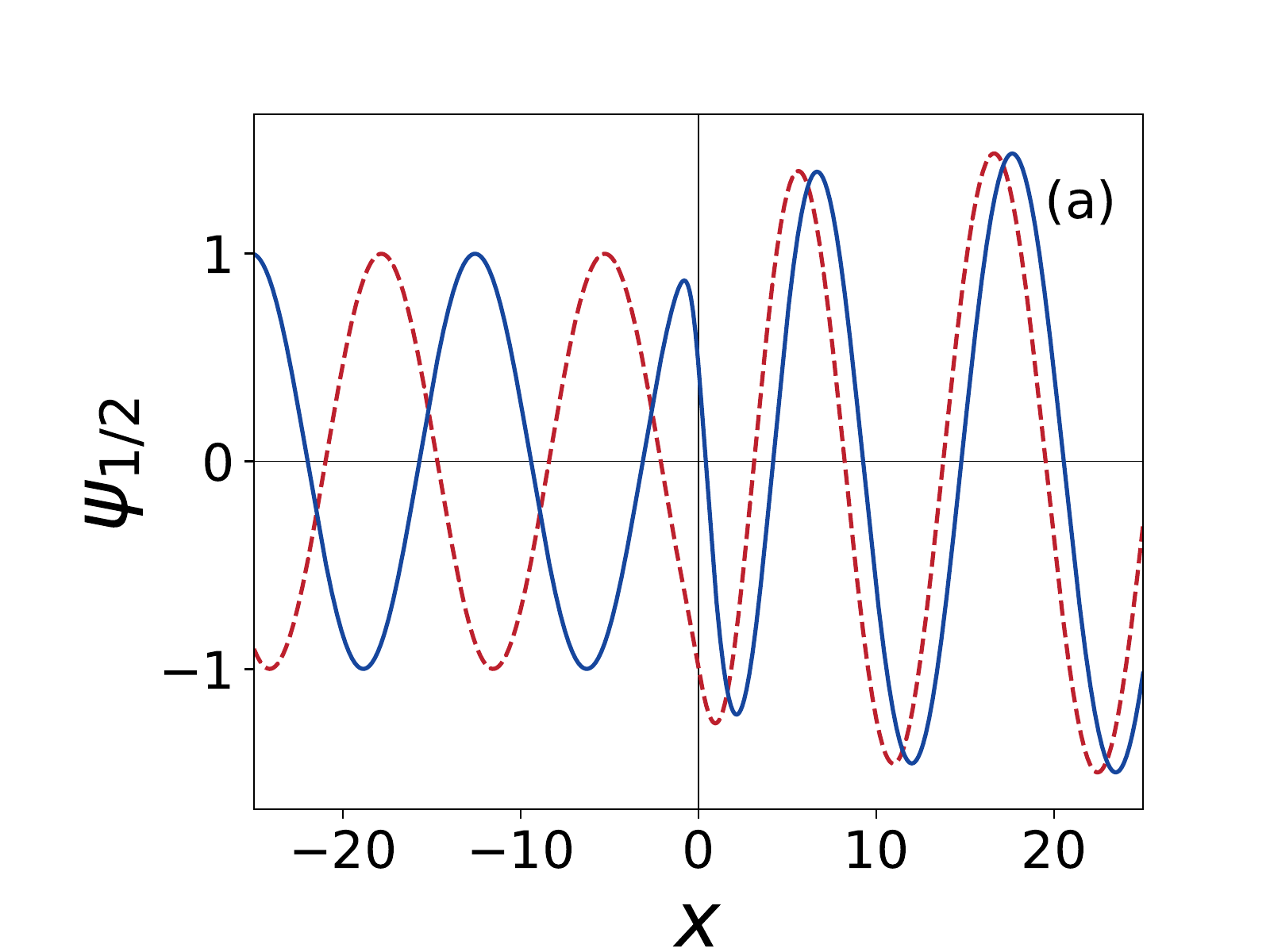}&
\includegraphics[width=0.45\textwidth]{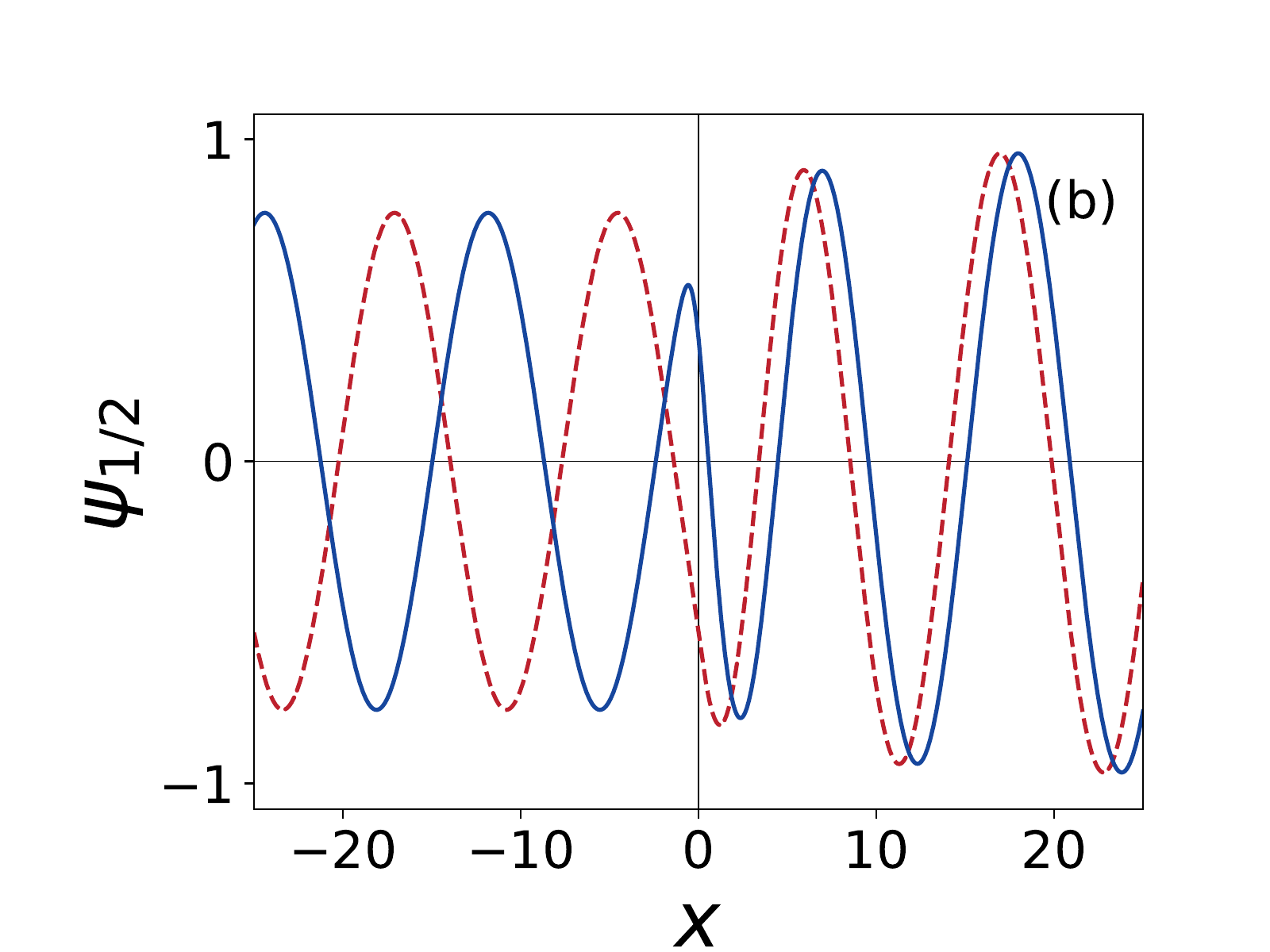}
\end{tabular}
\caption{(a) Fermionic continuum states in the case of the scattering from the right. (b) Fermionic continuum states in the case of the scattering from the left. In both graphs $g=0.9$ and $k=0.5$ and also the solid curve (blue) and dashed (red) curve show the upper and lower components, $\psi_1$ and $\psi_2$, respectively.}
\label{fig:fermionscatt}
\end{figure}

Fermions can be coupled to the soliton in various ways. We introduce a fermion field $\psi$ coupled to the soliton through a Yukawa coupling in the following form
\begin{align}
	\mathcal L = \frac{1}{2}\partial_\mu\phi\,\partial^\mu\phi - V(\phi) + \bar\psi i\gamma^\mu\partial_\mu\psi - g\,\phi\,\bar\psi\psi\,,
\end{align}
where $g$ is a coupling constant. The resulting equation of motion in the background of the soliton reads
\begin{align}
	i\gamma^\mu\partial_\mu\psi - g\,\phi\,\psi = 0.
\end{align}
Opening the spinor field $\psi$ in components as $\psi=e^{-iEt}\begin{pmatrix}\psi_1\\\psi_2\end{pmatrix}$ one can find the pair of equations
\begin{align}
	E\,\psi_1 + \psi_2'- g\,\phi\,\psi_2 &= 0\,,\nonumber\\
	E\,\psi_2 - \psi_1'- g\,\phi\,\psi_1 &= 0\,, \label{eqofmotion}
\end{align}
where the representation for the Dirac matrices is taken as $\gamma^0=\sigma_1$, $\gamma^1=i\sigma_3$ and $\gamma^5=\sigma_2$. For the case of $\phi^4$ model, a zero energy bound state or zero mode is known to exist, which is also the case for our model. The zero mode is given by
\begin{align}
	\psi(x) = \mathcal{N}
	\begin{pmatrix}
	  e^{-g\int^x\phi(x')\mathrm{d}x'}\\
	  0
	\end{pmatrix},
\end{align}
where $\mathcal{N}$ is the normalization constant. Since one of the components is null the soliton never receives backreaction from this state and the solution is exact \cite{AmadoMohammadi2017}.
Performing the above integration we can obtain an explicit solution to the state
\begin{align}
	\psi_1&=\mathcal{N}\exp\left\{{-g\int^x\left[1-\frac{2}{1+W\left[e^{1+4x'}\right]}\right]\mathrm{d}x'}\right\}\, .
\end{align}
Performing the change of variables $y=\exp(1+4x)$ this becomes
\begin{align}
	\psi_1 &= \mathcal{N}\exp\left\{{-g\,x +\frac{g}{2} \int^{e^{1+4x}}\frac{W'[y]}{W[y]}\mathrm{d}y}\right\}\,,
\end{align}
using the property of Lambert $W$ function 
\begin{align}
	W'[y] = \frac{W[y]}{y\left(1+W[y]\right)}\,.
\end{align}
Therefore, the wavefunction becomes
\begin{align} \label{normalzeromode}
	\psi &= \mathcal{N}\begin{pmatrix}\exp\left\{-g\,x +\frac{g}{2} \ln\left[W\left[e^{1+4x}\right]\right]\right\}\\0\end{pmatrix} \\&= \mathcal{N}\,\begin{pmatrix}\left(W\left[e^{1+4x}\right]\right)^{\frac{g}{2}}e^{-g\,x}\\0\end{pmatrix}\,,
\end{align}
with the normalization constant
\begin{align}
	\mathcal{N} = \sqrt{\left(\frac{g}{2}\right)^{g/2}\frac{2\,e^{-g/2}}{\Gamma\left[g/2\right]}}\,.
\end{align}
Details of the calculation above are supplied in appendix~\ref{app:N2}. In Fig.~\ref{fig:fermionzeromode}, we show the fermionic zero mode for two different values of the coupling. The resulting parity asymmetry in the zero mode of our model is visible, especially when the fermion-soliton coupling $g$ increases. 
\begin{figure}[htp]
\centering
\includegraphics[width=0.56\textwidth]{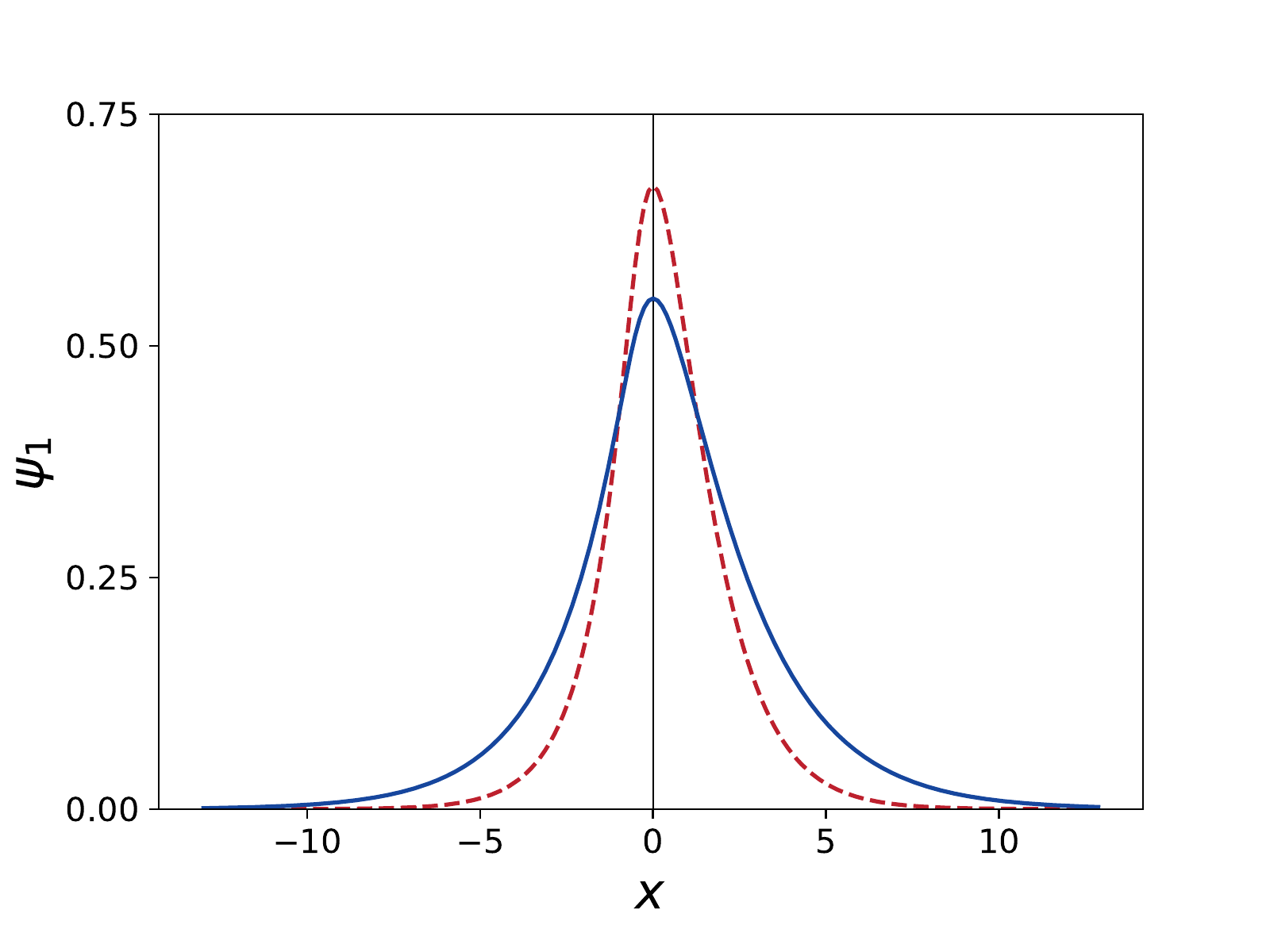}
\caption{Fermionic zero mode for $g=0.5$ (solid curve in blue) and $g=0.9$ (dashed curve in red).}
\label{fig:fermionzeromode}
\end{figure}
Besides that, we solve the equations of motion in (\ref{eqofmotion}) for nonzero bound states numerically where the result for the upper and lower components of the first and second fermionic bound states is presented in Fig.~\ref{fig:fermionbound}. A close inspection of the figure reveals that the states do not respect parity. Moreover, we plot the bound and threshold energies as a function of the bound state number as well as the fermion-soliton coupling $g$ in Fig.~\ref{fig:fermionspectrum}, where the system is solved numerically.  It is not difficult to show that the system has energy-reflection symmetry, which is given by $\gamma^1$ in our model. 
In Fig.~\ref{fig:fermionspectrum} the symmetry manifests itself by the symmetric form of the spectrum around $E=0$ line. 
For very small values of $g$, the only discrete mode is the zero mode. However, gradually increasing $g$ from zero supports more and more bound states.
Besides the bound states, one can explore the scattering ones considering energies above the threshold in the equation of motion (\ref{eqofmotion}). We show the upper and lower components of the fermionic scattering states for the scattering from both directions in Fig.~\ref{fig:fermionscatt}. Again, it is easy to observe that the states do not respect parity symmetry.

\section{Conclusion}

In this work, we have designed a parity-breaking solitonic model where the potential is up to sixth order in the scalar field $\phi$, with two minima. The soliton solutions connecting the two minima in a nonsymmetric form, with one long-range power-law tail and one exponential asymptotics, has been solved in terms of the Lambert W function. Although the system lacks $Z_2$ symmetry, changing $\phi \to -\phi$ only swaps the role of the soliton and antisoliton solutions. We have found the soliton mass, which is equal to the one for the kink of $\phi^4$ theory, despite a very different energy density. Studying the linear stability equation for the small perturbations around the static soliton solutions, we have concluded that the only discrete mode is the zero mode associated with the translational invariance, in contrast with the parity-symmetric $\phi^4$ model. 
Besides that, we have studied the interaction of the boson and fermion fields with the soliton considering two different types of interaction terms for the bosonic one and the Yukawa interaction for the fermionic one. The first interaction we have examined has led to a non-homogeneous Klein-Gordon equation with interesting results. For example, we have shown that the boson bound state also acquires the form of a defect, which means that the soliton in our model traps the bosonic field in a kink configuration. Considering the second interaction term, a Yukawa-like interaction, we have shown that one can write the equation of motion in the form of a Schr\"odinger equation. With a change of variables and mapping the parameters with the results obtained in \cite{ishkhanyan2016lambert}, we have found the bound and continuum states analytically. We have also studied the scattering of the waves from the left and right as well as the reflection coefficient, knowing the barrier shape potential term. We have shown that the reflection coefficient is unity for waves with energies beneath the barrier and goes to zero at high-energy, as expected.
We have solved the system analytically for both types of interactions. 
It is not common to find systems that can be fully solved analytically, and this makes the model more valuable for follow-up studies and applications.
Moreover, in both cases, we have verified that the results match the expectations in the limiting cases where $\phi(x\to\pm \infty)\to \pm 1$. Finally, the interaction of the fermion field with the soliton has been considered. In this case, we have been able to find the normalized fermion zero mode analytically. We have also obtained the nonzero bound energy spectrum as a function of bound state number as well as the fermion-soliton coupling $g$ numerically. The system has energy reflection symmetry given by $\gamma^1$ resulting in a symmetric bound energy spectrum.
We have shown that, for very small values of the coupling, the only discrete mode is the zero mode, with a growing number of bound states appearing as we gradually increase the coupling. Finally, the scattering oscillating modes for the waves coming from the left and the right have been shown.

In future work, we plan to apply the model proposed here to address the soliton-soliton long-range interactions, taking advantage of the analytical properties of the model.

\section*{Acknowledgments}
AA would like to thank CAPES for financial support under the PNPD fellowship. AM acknowledges the financial support from CNPq (process number 305893/2017-3), CAPES and Universidade Federal de Pernambuco Edital Qualis A. AA and AM are thankful to Carlos Batista for fruitful discussions.

\appendix
\section{Calculation of the soliton solution}\label{app:p2}
Performing the change of variables $\chi=\phi-1$ in eq.~\ref{phix} we obtain
\begin{align}
\log\left[\frac{1+2/\chi}{e^{2/\chi}}\right] = 4x -i \pi +2\,,
\end{align}
which leads to
\begin{align}
	\left(-1-2/\chi\right)e^{-1-2/\chi} = e^{4x+1}\,.
\end{align}
Recalling that the Lambert $W$ function is defined as the inverse function of 
\begin{align}
	f(W) = We^W\,,
\end{align}
the previous equation can be written as 
\begin{align}
	-1-2/\chi = W[e^{4x+1}]\,.
\end{align}
Now solving for $\chi$ and reintroducing $\phi$ we have the final result
\begin{align}
	\phi = 1-\frac{2}{1+W\left[e^{4x+1}\right]}\,.
\end{align}

\section{Integration constants of the bound states in model~I}\label{app:model1}
We start with the general form of the solution for model~I, eq.~(\ref{Model1Bound}),
\begin{align}\label{eq:app:model1}
	\chi_s(x) =&\, A e^{4 n x} + B e^{-4 n x}
	+\frac{g}{16n}\left[f_n(x) - f_{-n}(x) - \frac{1}{n}\right]\,,
\end{align}
where
\begin{align}
	f_n(x)\equiv n^{n} e^{n(1+4x)}\,\Gamma\left(-n,n\, W\left[e^{1+4x}\right]\right)\,.
\end{align}
Let us first look at the limit $x\to+\infty$. At this limit, $f_n(x)$ becomes
\begin{align}
	f_n(x\to+\infty) &\approx n^{n} e^{n(1+4x)}\,\Gamma\left(-n,n\left[1+4x - \ln(1+4x)\right]\right)\nonumber\\
	&\approx n^{n} e^{n(1+4x)}\,\left[\frac{1}{4x} e^{-n(1+4x)} n^{-1 - n}\right] = \frac{1}{4nx}.
\end{align}
Therefore, 
\begin{align}
	\lim_{x\to+\infty} f_n(x) = 0,
\end{align}
and similarly
\begin{align}
	\lim_{x\to+\infty} f_{-n}(x) = 0.
\end{align}
Since the $A$ term diverges in this limit and there is no other term to compensate it, the constant $A$ should be set to zero.

Now considering the  limit $x\to-\infty$ we can determine the remaining constant $B$. Using the expansion of the Lambert function for small arguments we have
\begin{align}
	f_n(x\to-\infty) &\approx n^{n} e^{n(1+4x)}\,\Gamma\left(-n,n\, e^{1+4x}\right)\nonumber\\&\approx n^{n} e^{n(1+4x)}\,\frac{1}{n!}\left[\frac{e^{n\,e^{1+4x}}}{\left(n\,e^{1+4x}\right)^n}(n-1)!+(-1)^n\Gamma\left(0,n\,e^{1+4x}\right)\right]\nonumber\\
	&\approx n^{n} e^{n(1+4x)}\left\{\frac{e^{n\,e^{1+4x}}}{n\left(n\,e^{1+4x}\right)^n}+\frac{(-1)^n}{n!}\left[-\gamma-\ln\left(n\,e^{1+4x}\right)+n\,e^{1+4x}\right]\right\}\,.\nonumber
\end{align} 
In the limit $x\to-\infty$, we can ignore the second term and replace $e^{n\,e^{1+4x}}$ with 1, which gives
\begin{align}
	\lim_{x\to-\infty} f_n(x) = \frac{1}{n}\,.
\end{align}
We need to deal with $f_{-n}(x)$ differently. For this function, we have
\begin{align}
	f_{-n}(x\to-\infty) &\approx (-n)^{-n} e^{-n(1+4x)}\,\Gamma\left(n,-n\, e^{1+4x}\right)\nonumber\\
	&= (-n)^{-n} e^{-n(1+4x)}\,\left[\Gamma\left(n\right)-\gamma\left(n,-n\, e^{1+4x}\right)\right]\,,
\end{align}
where $\gamma(s,z)$ is the lower incomplete gamma function. Therefore, we obtain
\begin{align}
	f_{-n}(x\to-\infty) &\approx (-n)^{-n} e^{-n(1+4x)}\,\left[\Gamma\left(n\right)-\frac{\left(-n\, e^{1+4x}\right)^n}{n}\right] \\
	&= (-n)^{-n} e^{-n} \,\Gamma\left(n\right)e^{-4nx} - \frac{1}{n}\,.
\end{align}
The first term diverges at $x\to-\infty$ and should be cancelled by the $B$ term in the full solution. As a result, eq.~(\ref{eq:app:model1}) becomes
\begin{align}
	\chi_s(x) = \frac{g}{16n}\left[f_n(x) - f_{-n}(x) - \frac{1}{n} -(-n)^{-n} e^{-n} \,\Gamma\left(n\right) e^{-4 n x}\right]\,.
\end{align} 

\section{Normalization of the fermionic zero mode}\label{app:N2}
In eq.~(\ref{normalzeromode}), we can find the normalization factor in the following way
\begin{align}
	\mathcal{N}^2 &= 1\left/\int_{-\infty}^\infty \psi_1^*\psi_1\,\mathrm{d}x\right. = 1\left/\int_{-\infty}^\infty e^{-2g\,x}\,\left(W\left[e^{1+4x}\right]\right)^g\,dx\right. \, .
\end{align}
Choosing the transformation $y=e^{1+4x}$, it results in
\begin{align}
	\mathcal{N}^2 &= 1\left/\int_0^\infty \frac{{e^{-\frac{g}{2}\,\left[\ln\left(y\right)-1\right]}}}{4\,y}\,\left(W\left[y\right]\right)^g \, dy\right. \\
	&= e^{-g/2}\left/\int_0^\infty \frac{1}{4\,y^{1+g/2}}\,\left(W\left[y\right]\right)^g \, dy\right.\,.
\end{align}
Now let's consider the change of variables $w=W(y)$ (notice that $y=we^w$, by the very definition of the Lambert $W$ function). Therefore,
\begin{align}
	\mathcal{N}^2 &= 4e^{-g/2}\left/\int_0^\infty \frac{w^g}{\left(we^w\right)^{1+g/2}}\,\left(1+w\right)e^w \, dw\right.\, ,
\end{align}
which leads to
\begin{align}
	\mathcal{N}^2 &= \left(\frac{g}{2}\right)^{g/2}\frac{2\,e^{-g/2}}{\Gamma\left[g/2\right]} \, .
\end{align}

\end{document}